\documentclass[12pt]{article}
\usepackage{amsmath}
\usepackage{graphicx}
\usepackage{enumerate}
\usepackage{apacite}
\usepackage{natbib}
\usepackage{url} 
\usepackage{amsgen,amsmath,amstext,amsbsy,amsopn,amssymb}
\usepackage{float}
\RequirePackage{bm}
\usepackage{graphicx}
\usepackage{tikz}
\usetikzlibrary{shapes}

\usepackage{hyperref}  
\usepackage{subfigure}
\usepackage[toc,page]{appendix}
\newcommand{\blind}{0}

\begin{document}

\def\spacingset#1{\renewcommand{\baselinestretch}%
{#1}\small\normalsize} \spacingset{2}


\if0\blind{ 
\title{\bf Measure of Strength of Evidence for Visually Observed Differences between Subpopulations}
  \author{Xi Yang\\
  Department of Statistics and Operations Research\\
University of North Carolina
Chapel Hill, USA\\
yangximath@gmail.com
\\
    Jan Hannig \\
    Department of Statistics and Operations Research\\
University of North Carolina
Chapel Hill, USA\\
jan.hannig@unc.edu
\\
Katherine A. Hoadley \\
    Department of Genetics\\
University of North Carolina
Chapel Hill, USA\\
hoadley@med.unc.edu\\
Iain Carmichael\\
Department of Statistics\\ University of California  Berkeley, USA\\
iain@berkeley.edu\\
J.S. Marron\\
    Department of Statistics and Operations Research\\
University of North Carolina
Chapel Hill, USA\\
marron@unc.edu}
  \maketitle
}\fi
\if1\blind{
  \bigskip
  \bigskip
  \bigskip
  \begin{center}
    {\LARGE\bf Measure of Strength of Evidence for Visually Observed Differences between Subpopulations}
\end{center}
  \medskip
}\fi

\bigskip
\begin{abstract}

For measuring the strength of visually-observed subpopulation differences, the Population Difference Criterion is proposed to assess the statistical significance of visually observed subpopulation differences. It addresses the following challenges: in high-dimensional contexts, distributional models can be dubious; in high-signal contexts, conventional permutation tests give poor pairwise comparisons. We also make two other contributions:  Based on a careful analysis we find that a balanced permutation approach is more powerful in high-signal contexts than conventional permutations. Another contribution is the quantification of uncertainty due to permutation variation via a bootstrap confidence interval. The practical usefulness of these ideas is illustrated in the comparison of subpopulations of modern cancer data.

\end{abstract}

\noindent%

\section{Introduction}
\label{intro}
In the age of Big Data, many contexts involve analyzing and understanding
relationships between multiple subpopulations. A fascinating and particularly
deep example of this comes from cancer research as illustrated in
Section \ref{subsec:Cancer-Data-Illustration}, where we have gene expression data from five cancer types and a goal of understanding relationships between the cancer types (aka subpopulations). A common approach to
understanding subpopulation differences is visualization using Principal Component Analysis (PCA) \citet{jolliffe1986principal}. While this method gives useful insights, visual approaches can be deceptive. This motivates our work to develop a method for quantification of the strength of the evidence for distinction between each pair of groups. 
Because of the very rich general structure of biological processes, classical
statistical distributional models are woefully inadequate. Hence permutation
testing methods, such as the DiProPerm test proposed by \citet{wei2016direction}, provide an appealing alternative. 

The DiProPerm method measures strength of difference between any two subpopulations
by projecting the data onto a direction aimed at separating them,
and summarizing using the difference of the projected means as a statistic.
Typical permutation p-values are calculated using the number of permuted
summaries that lie outside the corresponding true data summary statistic.
Bioinformatics data is frequently \emph{high signal}, meaning there are usually
very strong differences between subpopulations.
The challenge of high signal situations is that there frequently are
no permuted statistics outside that range, so the only conclusion
is that the permutation p-value is $<\frac{1}{n_{P}}$, where $n_{P}$
is the number of permutations. From the classical viewpoint, that is strong evidence
of all such differences being significant which is the end of the story.
However, understanding the relationship between subpopulations is
a different challenge.  
This motivates our invention of the \emph{Population Difference Criterion}
(PDC), which is a quantitative measure of separation between subpopulations
that provides meaningful comparisons even in high dimensional and
high signal contexts. 
A very important example of the value of PDC is in comparison
of the many possible pre-processing operations that are routinely
used for example in bioinformatics applications. In particular, better pre-processing
methods are those which result in a larger PDC for a given set of subpopulations
of interest. 

A major contribution of this paper discussed in Section~\ref{pdc-define} is the discovery of a peculiar phenomenon that in high signal situations increasing signal strength can actually entail the loss of statistical power as measured by the PDC. Detailed mathematical analysis reveals that this is caused by the traditional permutation scheme employed in DiProPerm \cite{wei2016direction}. This motivates our proposal of a non-standard {\em balanced permutation} scheme. Comparisons using simulated and real data show that balanced permutations provide more powerful results.

Yet another contribution of this paper appears in Section \ref{ci:all}, where we propose confidence intervals that account for the Monte Carlo uncertainty in permutation testing. The value of quantifying that uncertainty for comparing multiple cancer subpopulations is demonstrated in Section \ref{subsec:Cancer-Data-Illustration}. Discussion of controversies related to non-traditional permutation schemes can be found in Section~\ref{sb}.


{
\section{Population Difference Criterion}
\label{pdc-define}
Consider data from two potentially high-dimensional populations $X_1,\ldots,X_m$ and $Y_1,\ldots Y_n$. As explained in the introduction, a common way of visualizing the difference between the subpopulations uses projections on a given direction determined by a unit vector $w$, e.g., a PCA direction.
A particularly useful visual direction for distinguishing subpopulations is the Distance Weighted Discrimination (DWD)
direction vector $w$ proposed by \citet{marron2007distance}. 
{DWD solves an optimization problem that is formally stated in Appendix~\ref{ap:DWD}}. A potential drawback is that DWD can be relatively slow to compute. Therefore, we will also investigate a computationally faster version of DiProPerm based on projection onto the Mean Difference (MD) direction, i.e. $w=\frac{\bar X_m-\bar Y_n}{\|\bar X_m-\bar Y_n\|}$ -- the direction pointing from the mean of one group to the mean of the other.

While such visualizations are suggestive, they can also be deceptive.  As noted in \citet{wei2016direction}, rigorous quantification of visual differences can be surprisingly counter intuitive, because human intuition is not good at incorporating issues such as sample size and high dimensional variation into visual impression.
 An explicit example of this is shown in Figure 2.3 of the PhD dissertation of \cite{yang2021machine}. Hence it is very important to provide a quantitative measure of the strength of the evidence for visual subpopulation separation  based on rigorous statistical inference.

An early version of a quantitative measure  of 
this type was provided by the DiProPerm test (\citet{wei2016direction}).
The DiProPerm test statistic is the observed mean difference of the projected scores:
\begin{equation}\label{eq:Cdef}
C = \left|\frac 1m\sum_{j=1}^m \left<w, X_k\right> - \frac 1n\sum_{k=1}^n \left<w,Y_k\right>\right|,
\end{equation}
where $\left<w,x\right>$ is the inner product.

The \emph{Population Difference Criterion} (PDC) is a quantitative measure of the differences between subpopulations that may be visually apparent in PCA scatter plots. Specifically, 
\[
PDC=\frac{C-E{C}}{\sqrt{\operatorname{Var} C}},
\]
where the mean and variance of $C$ are computed under the null model of no difference between subpopulations.
Larger values of PDC represent stronger evidence of the difference between subpopulations. 

Traditionally, the null distribution $EC$ and $\operatorname{Var} C$ have been estimated using permutation based methods. In particular,
\begin{equation}\label{eq:PDC}
\widehat{PDC}=\frac{C-\bar C}{S},
\end{equation}
where the mean $\bar C$ and standard deviation $S$ are estimated using re-sampling of the class labels and re-projection of the data.
Note that if the null distribution of $C$ was Gaussian, the PDC would be the classical Z-score and was called that in  \citet{wei2016direction}. But that term is not used here as the null distributions of $C$ could be far from Gaussian.
}

{
\subsection{Gaussian model}
\label{cbp}
In this section, we study the behavior of the DiProPerm PDC using a basic two-class Gaussian model. Both classes are assumed to follow the multivariate normal distribution with means separated by $2g$, i.e., 
  \begin{equation}\label{eq:SimModel}
 \begin{gathered}
  \mbox{Class +1 } (X):\qquad   X_{j}\sim N_d(g\cdot u,\sigma^2 I),\quad j=1,...,m\\
 \mbox{Class -1 } (Y):\qquad  Y_{k}\sim N_d(-g\cdot u,\sigma^2 I),\quad k=1,...,n
  \end{gathered}
  \end{equation}
where $u\in \mathbb{R}^d$ is some unit vector, such as $[1/\sqrt{d} \cdots 1/\sqrt{d}]^T$ or $[1\ 0 \cdots 0]^T$, and $\sigma \in \mathbb{R}^+$. The actual direction of $u$ is irrelevant because the DiProPerm test is rotation invariant.

An important component of our mathematical analysis is the derivation of the distribution of the observed statistics \eqref{eq:Cdef} under the model \eqref{eq:SimModel}. Using the MD direction
\[
 C^2=\|\bar X-\bar Y\|^2=\sum_{k=1}^d(\bar X^k-\bar Y^k)^2\sim (\frac{1}{m}+\frac{1}{n})\sigma^2 \chi^2_d(\frac{4g^2}{ (\frac{1}{m}+\frac{1}{n})})
\]
and so
\begin{equation}\label{eq:chi}
C=\|\bar X-\bar Y\|\sim \sigma \sqrt{\frac{1}{m}+\frac{1}{n}}\chi_d(\sqrt{\frac{4g^2}{ (\frac{1}{m}+\frac{1}{n})}}).
\end{equation}
Here $\chi_d(\lambda)$ is called the \textit{non-central Chi distribution} \citep{johnson1972continuous},  the square root of the non-central Chi-square distribution with degrees of freedom $d$ and non-centrality parameter $\lambda^2$.

Under the null hypothesis $g=0$ we have $EC=\sigma \sqrt{\frac{1}{m}+\frac{1}{n}}\sqrt{2}\frac{\Gamma((d+1)/2)}{\Gamma(d/2)}$ and $\operatorname{Var} C=\sigma^2(\frac{1}{m}+\frac{1}{n})(d-2(\frac{\Gamma((d+1)/2)}{\Gamma(d/2)})^2)$.
In general, 
\[
E(\chi_d(\lambda))
=\sqrt{\frac{\pi}{2}}\cdot L_{1/2}^{d/2-1}(\frac{-\lambda^2}{2}),
\]
where $ L_{n}^{a}(z)$ is the generalized Laguerre polynomial \citep{koekoek1993generalization}. Then we have
\begin{eqnarray}
\label{A}
E[\|\bar X-\bar Y\|]&=&\sigma \sqrt{\frac{1}{m}+\frac{1}{n}}\cdot\sqrt{\frac{\pi}{2}}\cdot L_{1/2}^{d/2-1}(-\frac{2g^2}{ (\frac{1}{m}+\frac{1}{n})})\\
Var[\|\bar X-\bar Y\|]&=&\sigma^2(\frac{1}{m}+\frac{1}{n})(d+\frac{4g^2}{ (\frac{1}{m}+\frac{1}{n})}-\frac{\pi}{2}\cdot( L_{1/2}^{d/2-1}(-\frac{2g^2}{ (\frac{1}{m}+\frac{1}{n})}))^2)
\end{eqnarray}
}

{
\subsection{Permutation distribution}
\label{Permutation distribution-f}
Since in practice, most people use the estimated PDC \eqref{eq:PDC}, it is important to study the behavior of the test statistic $C$ under various permutation schemes. 
The classical approach is to randomly reshuffle the class labels $\{-1,1\}$ for all observations. This generates {permuted data} $X_{j,per},\ j=1,\ldots,m$ and $Y_{k,per},\ k=1,\ldots,n$.  We call this approach {\em all permutations}. Let $C_i,\ i=1,\cdots, N$ be independent realizations of the test statistic \eqref{eq:Cdef} computed using permuted data. Again, in the case of the MD direction 
$
C_i=\|\bar X_{per} - \bar Y_{per}\|,\quad i=1, \cdots, N.
$ 

When labels are randomly reshuffled, there is a random number $R_i$ of observations in each class that switch labels. Note that when all permutations are used, $R_i$  is a Hypergeometric random variable whose probability mass function is:
\[
p_R(r)=P(R_i=r)=\frac{{m\choose r}{n\choose n-r}}{{m+n\choose m}},\quad r=0, 1, ..., \min(m,n). \label{rrrr}
\]

The conditional distribution of  $\bar X_{per}- \bar Y_{per}$ is (detailed derivation shown in  Appendix~\ref{a_condition}):
\begin{equation}
\label{gor}
(\bar X_{per}- \bar Y_{per})|R_i=r\sim N_d(2g\cdot (1-\frac{r}{m}-\frac{r}{n}) \cdot u, \sigma^2 (\frac{1}{m}+\frac{1}{n}) I_d).
\end{equation}
Thus, for a given permutation, $\frac{R_i}{m}$ is the proportion of the original Class -1 cases that are relabeled as the permuted Class +1, and $\frac{n-R_i}{n}$ is the proportion of the Class -1 cases that remain in the new Class -1. The difference between these 2 proportions, denoted as
\begin{equation}
\label{cou}
\xi_i=\xi(R_i)=\frac{n-R_i}{n}-\frac{R_i}{m}=1-\frac{R_i}{m}-\frac{R_i}{n},
\end{equation}
quantifies the class balance of this permutation. Hence we call it the \textit{coefficient of unbalance}. Those permutations with $\xi_i=0$ are  called \textit{balanced permutations}.

To study the theoretical behavior of MD PDC we need to calculate
the mean and variance of $\|\bar X_{per}-\bar Y_{per}\|$ under the model \eqref{eq:SimModel}. First
define the following notation:
\begin{equation*}
g_r=g\cdot \xi(r)=2g\cdot (1-\frac{r}{m}-\frac{r}{n}),\qquad
\lambda_r=\sqrt{\frac{4g_r^2}{\sigma^2(\frac{1}{m}+\frac{1}{n})}}.
\end{equation*} 
The unconditional distribution of $\bar X_{per}- \bar Y_{per}$ is a normal mixture which is related to Theorem 1 of \citet{wei2016direction}:
\[
\bar X_{per}- \bar Y_{per} \sim \sum_{r=0}^{\min(m, n)}p_R(r)N_d(2g_r\cdot u, \sigma^2 (\frac{1}{m}+\frac{1}{n}) I_d).
\]
The permutation null distribution is
\begin{equation}
\label{eq:mixture}
C_i=\|\bar X_{per}-\bar Y_{per}\|\sim\sigma \sqrt{\frac{1}{m}+\frac{1}{n}}\sum_{r=0}^{\min(m, n)}p_R(r) \chi_d(\lambda_r).
\end{equation} 
Thus $C_i$ has a mixture distribution with the mixture component driven by the coefficient of unbalance $\xi_i$.
Then we have
\begin{eqnarray}
\label{B}
E(\|\bar X_{per}-\bar Y_{per}\|)&=&\sigma \sqrt{\frac{\pi}{2}}\cdot\sqrt{\frac{1}{m}+\frac{1}{n}}\sum_{r=0}^mp_R(r)  L_{1/2}^{d/2-1}(\frac{-\lambda_r^2}{2})\\
\label{V}
Var(\|\bar X_{per}-\bar Y_{per}\|)&=& \sigma^2 (\frac{1}{m}+\frac{1}{n})\sum_{r=0}^mp_R(r) (d+\lambda_r^2)- E(\|\bar X_{per}-\bar Y_{per}\|)^2
\end{eqnarray}
Note that, the sample sizes $m, n$ are inversely related to the sample variances. 

Using \eqref{A}, \eqref{B}, \eqref{V} define the PDC function
\begin{equation}\label{eq:PDCexpF}
f(m, n, d, g) := \frac{E(C)-E(C_i)}{\sqrt{Var(C_i)}}=\frac{E[\|\bar X-\bar Y\|]-E(\|\bar X_{per}-\bar Y_{per}\|)}{\sqrt{Var(\|\bar X_{per}-\bar Y_{per}\|)}}.
\end{equation}
This function provides important lessons about the behavior of the all permutation based PDC under the Gaussian model.
In particular, careful examination of $f(m, n, d, g)$ viewed as a function of $g$ (with $m,n,d$ fixed) shows that the function first increases and then decreases, eventually converging to
\begin{eqnarray}
\label{f}
\lim_{g\to \infty}f(m, n, d, g)=\frac{1-\sum_{r=0}^m  p_R(r)  |1-r/m-r/n|}{\sqrt{\frac{1}{m+n-1}-(\sum_{r=0}^m  p_R(r)  |1-r/m-r/n|)^2}}.
\end{eqnarray}
(see Appendix~\ref{ap:limit}). This behavior is a serious deficiency of the traditional all permutation approach since the power of the test fails to increase with stronger signal.
}

{
\subsection{Gaussian model simulation}
\label{Gaussian model simulation}
In this section, we demonstrate the behavior of the DiProPerm PDC using a simulation based on the Gaussian model \eqref{eq:SimModel}.  
In each of our simulation scenarios we generate samples from the two populations of size $m=n=100$. We consider three very different dimensions $d=1,10,100,$ and 200 values of signal strength $g$ ranging equally between 0 and 20. The PDC values \eqref{eq:PDC} are calculated using the mean and variance of $N=100$ permutations. Each $d,g$ combination is replicated 100 times for the PDC calculated using MD and only 10 times using DWD due to the longer running time of DWD.

The results are summarized in Figure~\ref{011}. The left and right panels show the PDC calculated using the MD and DWD directions respectively. In both panels the horizontal axis shows the signal strength $g$ and the vertical axis shows the DiProPerm PDC. The solid curves in both panels show local linear regression  estimates \citep{FanGijbels1996} of the PDC samples vs signal strength $g$. The dashed curve on the left shows the theoretical PDC value $f(m,n,d,g)$ viewed as the function of $g$ computed in \eqref{eq:PDCexpF}.
Each dot is a single realization of the estimated PDC value, reflecting its variation due to randomness. 

When using the MD direction and $d>1$ (left panel), the PDC first goes up (as expected from increasing signal strength), then goes down (which is quite surprising), and finally converges to the limit given by \eqref{f}. When using the DWD direction the PDC also levels off instead of increasing, as one would na\"ively expect, with increasing signal strength. 
\begin{figure}[H]
  \centering
  \includegraphics[width=15cm]{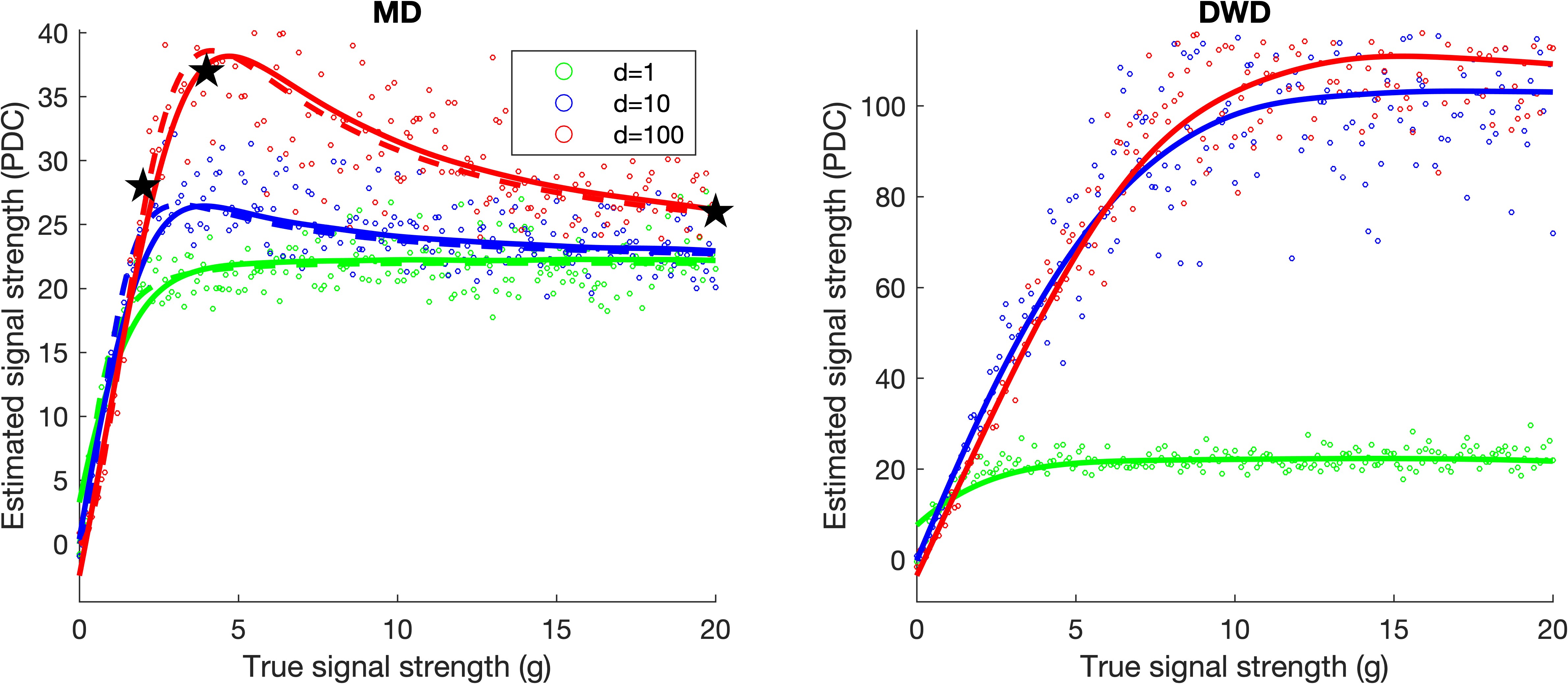}\\
   \caption{Test power as indicated by PDC, with MD on the left and DWD on the right, for different choices of $d$ (shown with colors) and signal strength $g$ (x-axes).   The y-axes show the PDC from DiProPerm's results. The dashed curves in the left panel are the theoretical PDC \eqref{eq:PDCexpF}. The solid curves in both panels are local linear regression fits of the Monte Carlo samples. Each dot is a single realization of the estimated PDC value. Three representative cases studied in Figure~\ref{012} are highlighted in the left panel as black stars. We observe that PDC does not increase with signal strength $g$ as expected.}
  \label{011}
\end{figure}

The non-intuitive behavior of PDC computed using all permutations is further explored in Figure~\ref{012} by taking a deeper look at three particular data sets with $d=100$, $g=2,4,20$. These 3 values of $g$ represent increasing, peak, and decreasing regions of the PDC computed using the MD direction. They are highlighted as black stars in Figure~\ref{011} and correspond to the  3 columns in Figure~\ref{012}. The permuted distributions are shown in the top three panels of Figure~\ref{012} using a jitter plot (\citet{tukey1976exploratory}), where on the horizontal axis we plot the test statistic $C_i$, the independent realizations of the test statistic \eqref{eq:Cdef} computed using permuted data and the MD and DWD direction respectively, and on the vertical axis we plot a random height for visual separation.  The colors of the dots in these 3 top panels represent the absolute value of coefficient of unbalance $|\xi_i|$ of the $i$th permutation, using the color bar shown in Figure~\ref{000}.  The black curves are kernel density estimates (KDE, i.e., a smooth histogram, \citet{wand1994kernel}) of the distribution of the permuted test statistics (dots). The numbers on the vertical axes are the height of the kernel density estimate.
\begin{figure}[H]
  \centering
  \includegraphics[width=15cm]{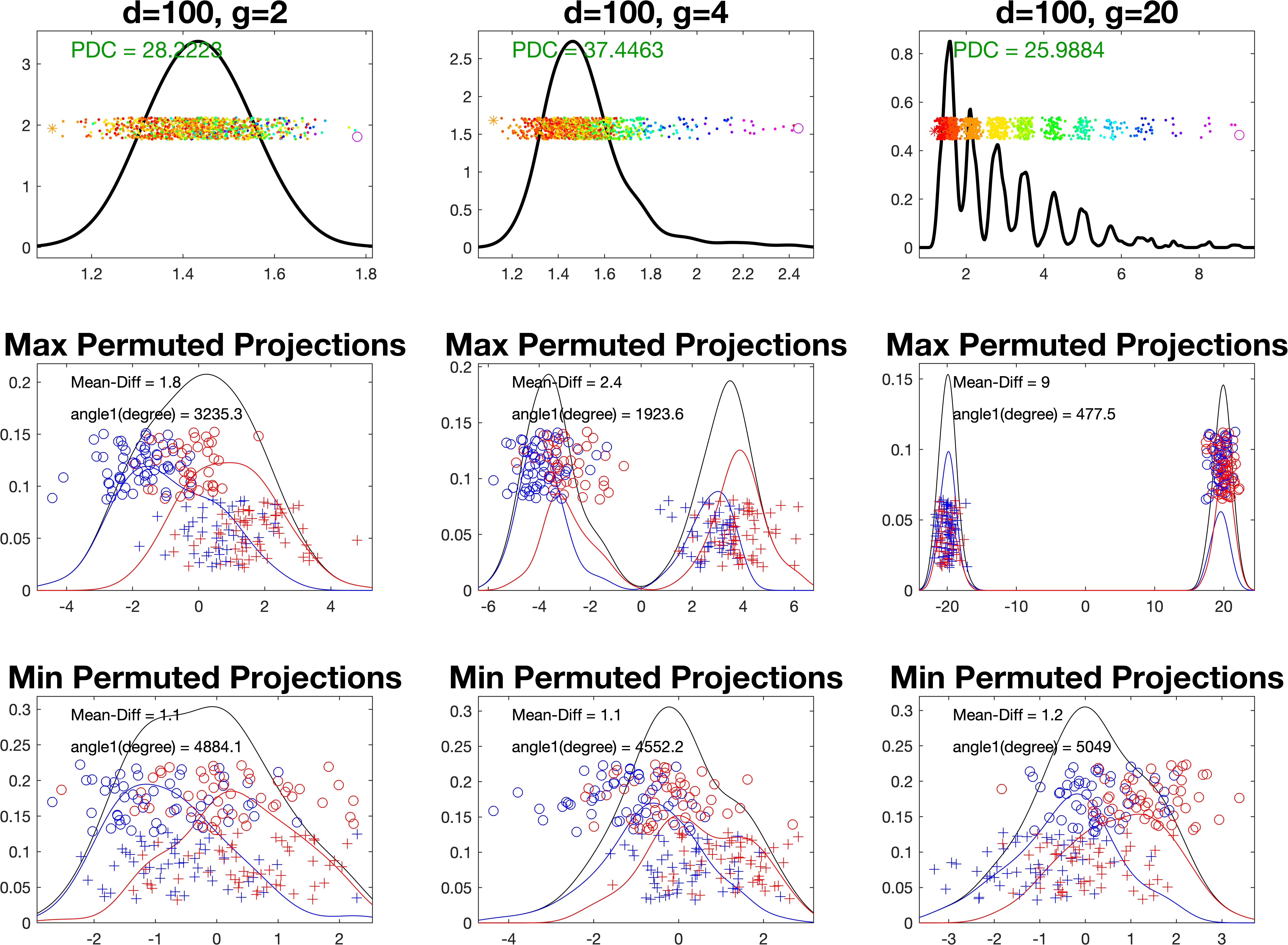}\\
  \caption{DiProPerm results for the MD direction and $d=100$. Left: $g=2$; middle: $g=4$; right: $g=20$ shown as black stars in Figure~\ref{011}. Top panels are the permuted statistics $C_i$ and their kernel density estimates with PDCs values printed in green text. Middle and bottom panels are chosen permutations with colors representing the permuted labels and symbols representing the original labels; the middle row of panels have the largest permuted statistic (colored circles in the top panels); bottom panels are permutations with the smallest permuted statistic (colored stars in the top panels). Shows increasing skewness of the permutation distribution as $g$ grows due to permutation unbalance.}
  \label{012}
\end{figure}
\begin{figure}[H]
  \centering
  \includegraphics[width=15cm]{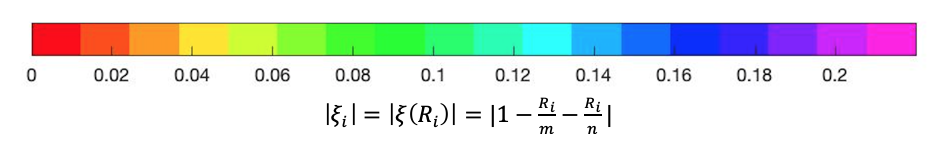}\\
  \caption{Colorbar used in the jitter plots in the top panels in Figure \ref{012}. Numbers represent the absolute value of the coefficient of unbalance $\xi_i$ defined in Equation \eqref{cou} in Section \ref{Permutation distribution-f}.}
  \label{000}
\end{figure}

From left to right, the kernel density estimates become more skewed and multi-modal in agreement with the fact that the all permutation null distributions are mixtures of chi-distributions \eqref{eq:mixture}. This is particularly apparent in the top right panel. As the signal, $g$, gets stronger there is much more separation of colors based on $|\xi_i|$. In the top left panel, which has the weakest signal ($g=2$), the colored dots are mostly mixed. In the top middle panel, as the signal strength ($g=4$) increases, the colored dots separate more. 

The multimodality is further explored in the bottom two rows which show the projections on the permuted directions for the smallest and largest $C_i$. In each case, the symbols represent the original class labels and the colors show the permuted labels, whose mean difference determines the direction. The $x$-axis is the projection scores on the permuted directions. The symbols in the middle are jitter plots and the heights of the symbols are random heights. The curves are kernel density estimates of the projection scores. The colors represent the permutated labels and symbols represent the original labels. Subdensities, corresponding to each permuted subpopulation, are shown using colors that correspond to the symbols. 
The $y$-axis shows the height of the KDE densities.

The middle row of Figure~\ref{012} shows the permutation with maximal $C_i$ for that $d$ and $g$ corresponding to the far-right circles in each top panel. Going from left to right, the permuted mean difference direction first separates the red/blue permuted class colors and then tends to separate the symbols (the original class labels). This direction essentially becomes the original mean difference direction of the non-permuted data for large $g$. This effect is usefully quantified by the angles between the observed mean difference and each permutation direction shown in each panel in the middle row. A large angle suggests a large discrepancy between the original mean difference and the corresponding permutation direction. The left panel $g=2$ is separating the colors well and mixing up the symbols with a relatively large angle $56^{\circ}$, as intuitively expected from the permutation test. This results in a PDC reflecting no signal as expected from the permutation test. In the middle $g=4$ panel, there is still some color separation but also a strong separation of the symbols, with a smaller angle, $33^{\circ}$. In the right $g=20$ panel, the angle is very small, $8^{\circ}$, showing this direction is very close to the mean difference direction of the original data. Because of the true class difference and the large coefficient of unbalance, this results in PDC values that are much larger than would be expected under the null distribution of no signal ($g=0$) which results in a strong loss of power. 

The bottom three panels of Figure~\ref{012} show the permutation with minimal $C_i$ corresponding to the far-left stars in each top panel.  However, because $|\xi_i| \approx 0$, the large $g$ doesn't affect these nearly balanced permutations as seen in the bottom panels, where the angles are relatively large and close to $90^{\circ}$ ($82^{\circ}, 79^{\circ}, 88^{\circ}$) as expected for random directions from the results of \citet{hall2005geometric}. 
Hence the bottom panels show projections which are much more consistent with the null hypothesis of no signal. 

\subsection{Proposed solution: Balanced Permutations}
\label{c}
As discussed above DiProPerm PDC has issues with power caused by the fact that the estimates of $EC$ and $\operatorname{Var} C$ under the null hypothesis is inflated when using the traditional all permutations. In this section, we propose a solution to this issue by using only balanced permutations. Recall that a permutation is called balanced if $\xi_i=1-R^i\times(1/m+1/n) \approx 0$, so the number of switches in labels is close to $\frac{mn}{m+n}$.

Figure~\ref{t} provides a simple example illustrating the difference between balanced and all permutations. There are 8 cases in each class shown as rows. The first row is colored using the real class labels followed by 7 permutations where symbols represent the true class labels and colors represent the permuted class labels as in the bottom 6 panels of Figure~\ref{012}. The colors of the text on the right are in the spirit of the color bar in Figure~\ref{000} and the colored dots in the top panels in Figure~\ref{012}. The top 3 permutations are all balanced permutations and in these cases solving the equation: $\xi_0=1-R^0\times(1/m+1/n)=0$ results in $R^0=\frac{mn}{m+n}=\frac{8\times 8}{8+8}=4$. The bottom 4 permutations are all unbalanced. The original DiProPerm draws from all permutations, but the proposed improved DiProPerm only draws from balanced permutations as shown in Figure~\ref{t}. 
\begin{figure}[H]
  \centering
  \includegraphics[width=15cm]{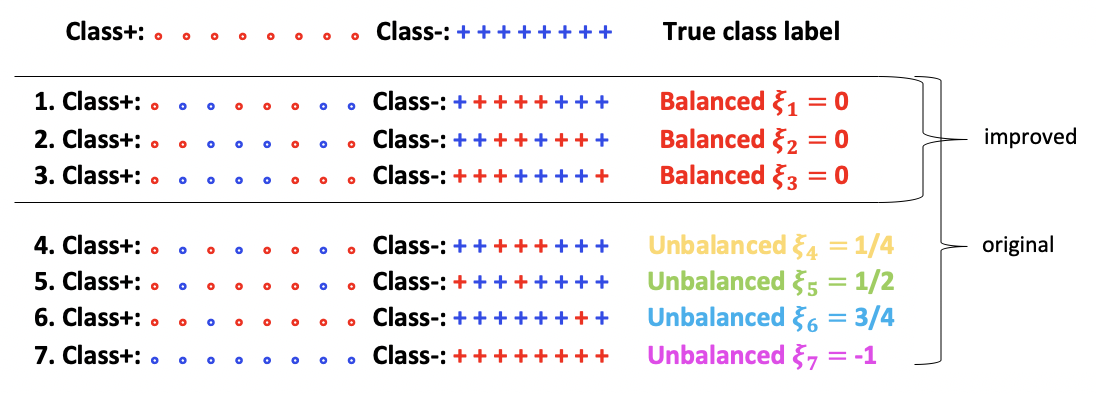}\\
  \caption{This figure shows 7 permutations from a simple example. The top row shows the true class labels and the rest of the rows show 7 different permutations represented as red and blue colored reassignments. The right column distinguishes between balanced and all permutations by coloring the text in the spirit of Figure~\ref{000}.} 
  \label{t}
\end{figure}

When there is a large separation between the centers of the two classes, using only the balanced permutations makes the mean of the permutation null distribution closer to zero. Therefore, it provides a more useful alternative distribution. In particular, under the Gaussian model \eqref{eq:SimModel}, Equation \eqref{gor} implies that for balanced permutations and the MD direction
 \[
 (\bar X_{per}- \bar Y_{per})\sim N_d(0, \sigma^2 (\frac{1}{m}+\frac{1}{n}) I_d).
 \]
Consequently,
\begin{eqnarray*}
E_{b}(\|\bar X_{per}-\bar Y_{per}\|)&=&\sqrt{2}\sigma \cdot\sqrt{\frac{1}{m}+\frac{1}{n}}\frac{\Gamma((d+1)/2)}{\Gamma(d/2)}\\
\operatorname{Var}_{b}(\|\bar X_{per}-\bar Y_{per}\|)&=& \sigma^2 (\frac{1}{m}+\frac{1}{n})[d-2(\frac{\Gamma((d+1)/2)}{\Gamma(d/2)})^2],
\end{eqnarray*}
which gives the balanced PDC curve as:
\begin{equation}
f_{b}(m, n, d, g):= \frac{E(C)-E_{b}(C_i)}{\sqrt{\operatorname{Var}_{b}(C_i)}}=\frac{E[\|\bar X-\bar Y\|]-E_{b}(\|\bar X_{per}-\bar Y_{per}\|)}{\sqrt{\operatorname{Var}_{b}(\|\bar X_{per}-\bar Y_{per}\|)}}.
\label{balancedf}
\end{equation}

Figure~\ref{81} compares PDCs computed using balanced vs. all permutations by adding the former to a part of Figure~\ref{011} for both MD and DWD directions. The lower curves labeled as {\em All permutations} are the same as the curves in Figure~\ref{011}. The higher level curves, labeled as {\em Balanced permutations}, use colors and symbols analogous to Figure~\ref{011} with \eqref{balancedf} replacing \eqref{eq:PDCexpF} for the dashed curve. Each dot is a single realization of the estimated balanced PDC value, reflecting its variation due to randomness.  The two sets of curves give direct comparison between the original and proposed versions of the DiProPerm PDC. For small $g$ the PDCs overlap. When the signal reaches a certain level, the balanced PDCs continue increasing as expected (from the increased signal strength), and the all permutation PDCs reach a peak and then seem to decrease (MD) or stay constant (DWD). This indicates the balanced PDC is much more powerful than the all permutation PDC in the case of strong signals for both MD and DWD directions. A careful look at the axis labels shows that stronger signals are required to see this effect for DWD.

\begin{figure}[H]
  \centering
  \includegraphics[width=15cm]{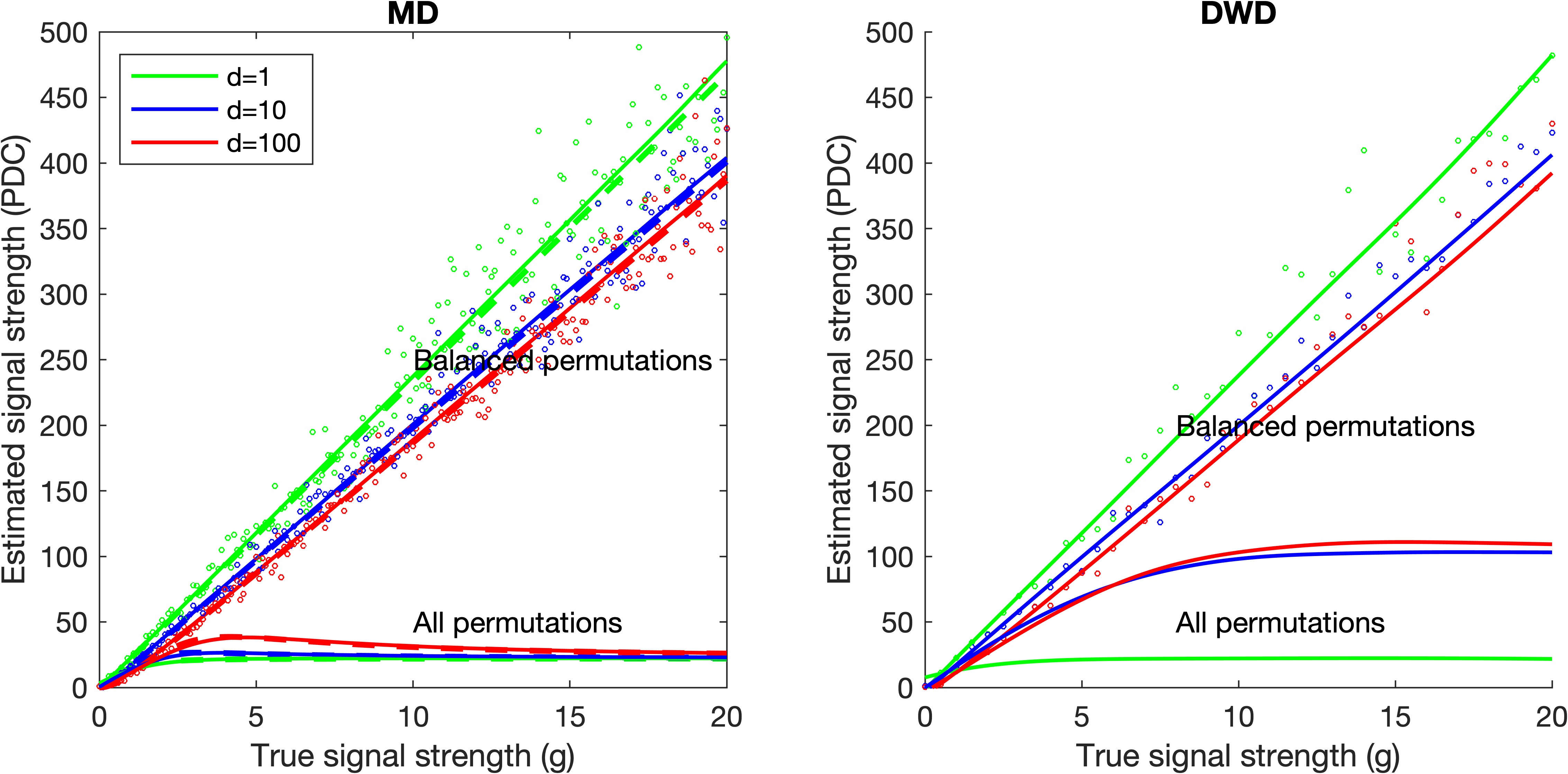}\\
    \caption{Realizations of the PDC for different choices of $d$ for both all and balanced permutations based on DiProPerm. The lower curves in both panels, labeled as {\em All permutations}, are the same curves as Figure~\ref{011}. The higher curves and dots in both panels, labeled as {\em Balanced permutations}, show the proposed DiProPerm using analogous colors and signs as in Figure~\ref{011}. Unlike the all permutation PDCs, the balanced permutation PDCs keep growing with larger signal strength $g$.}
    \label{81}
\end{figure}
}

Next, we continue the investigation of the three cases studied in Figure~\ref{012}. In Figure~\ref{3_densities}, the dashed curves are the derived theoretical null distribution using the MD direction, i.e. the scaled central $\chi_d$ distribution \eqref{eq:chi}. The goodness of fit of that distribution is demonstrated by the red solid curves which are kernel density estimates of the red dots in the jitter plots, i.e. only the balanced permutations. The solid black curves are the kernel estimates of all permutations. The bottom panels only show the red and black solid curves since the theoretical distribution using the DWD direction is much harder to derive. In both top and bottom panels, as the signal $g$ grows stronger, the distributions of all permutations (black curves) become more and more skewed but the distributions of balanced permutations (red dots/curves) stay the same and hence provide a much more useful null distribution. The skewness in the bottom panels is not as strong as in the top panels, indicating the DWD direction suffers less from the all permutations effect and has higher test power than using the MD direction. 
\begin{figure}[H]
  \centering
  \includegraphics[width=15cm]{picture/new_figure_08_13.jpg}\\
    \caption{Null distributions of the three representative testing contexts studied in Figure~\ref{012}. The MD and DWD directions are contrasted in the top and bottom panels. The dashed curves are the derived theoretical null distribution. The red solid curves are kernel density estimates of the balanced permutations (red dots in the jitter plots).  The black curves are the kernel density estimates of all permutations (dots of all colors). The red curves do not change with $g$ indicating a good estimate of the null distribution.}
    \label{3_densities}
\end{figure}

\section{Quantification of Permutation Sample Variation}\label{ci:all}
While they provide useful comparisons between data sets, the estimated permutation p-values and PDCs inherit variation caused by random sampling from each set of permutations. In some cases, this variation can obscure important differences between classes which motivates careful quantification of this uncertainty using confidence intervals.



The DiProPerm PDC \eqref{eq:PDC} is a random variable that depends on the permutation null distribution. Thus, a confidence interval for the PDC can be estimated by the upper and lower quantiles using bootstrap re-sampling methods. A general algorithm is based on $B$ repetitions. In our calculations $B=100$:
\begin{enumerate}
\item Draw a $B\times N$ matrix where each row is a random sample (with replacement) from the $N$ permutations used in the original calculation of PDC. Calculate the sample means and variances of each row. This results in $B$ re-sampled means and variances, which are used to get $B$ re-sampled PDCs. 
\item Find the upper and lower quantile of the PDCs based on the $B$ re-sampled PDCs in Step 2. 
\end{enumerate}
Note that this method is unrelated to the direction choices of DiProPerm, e.g., MD or DWD, and to the choice of balanced or all permutations.

Alternatively, as we discussed in Section~\ref{pdc-define} when the original data are close to normal, the permutation null distribution is a mixture the $\chi$ distributions \eqref{eq:mixture}. Thus, we can also estimate the distribution using the method of moments estimation based on the Welch–Satterthwaite approximation (\citet{satterthwaite1946approximate}). This has been explored in the Section 3 of the PhD Dissertation \cite{yang2021machine}. However, the normal assumption of the original data is often questionable, and hence the bootstrap re-sampling method is recommended.

\section{TCGA Pan-Can Data}
\label{subsec:Cancer-Data-Illustration}

To demonstrate the proposed method, we consider gene expression for five different cancer types, one of which is very different from the rest and two of which are similar to each other. Here, we used a subset of the TCGA Pan-Cancer data representing 1523 cases from 5 cancer types including 12478 genes. The tissues came from different organs (hence different cancer types), and represent a useful cohort to illustrate our proposed method for quantitatively determining their level of similarity or dissimilarity \citep{hoadley2018cell, hutter2018cancer}. These cancer types will be contrasted in visualizations discussed below using the colors and symbols shown in Table~\ref{t1}.
\begin{table}
\begin{tabular}{lllll}
Cancer & Abbreviation & Color & Symbol & Number \\
\hline
Acute Myeloid Leukemia & LAML & magenta & $\triangleleft$ & 173 \\
Bladder Urothelial Carcinoma  & BLCA & blue & *  & 138 \\
Breast Cancer  & BRCA & cyan & + &  950\\
Colon Adenocarcinoma  & COAD & yellow & $\bigstar$ & 190 \\
Rectal Adenocarcinoma& READ &red  & $\diamondsuit$ & 72
\end{tabular}
\caption{Abbreviated name, color, symbol (used in figures in this paper) and number of cases for each cancer type. Breast Cancer (BRCA) has the largest number of cases.
\label{t1}}
\end{table}

\begin{figure}
  \centering
  \includegraphics[width=10cm]{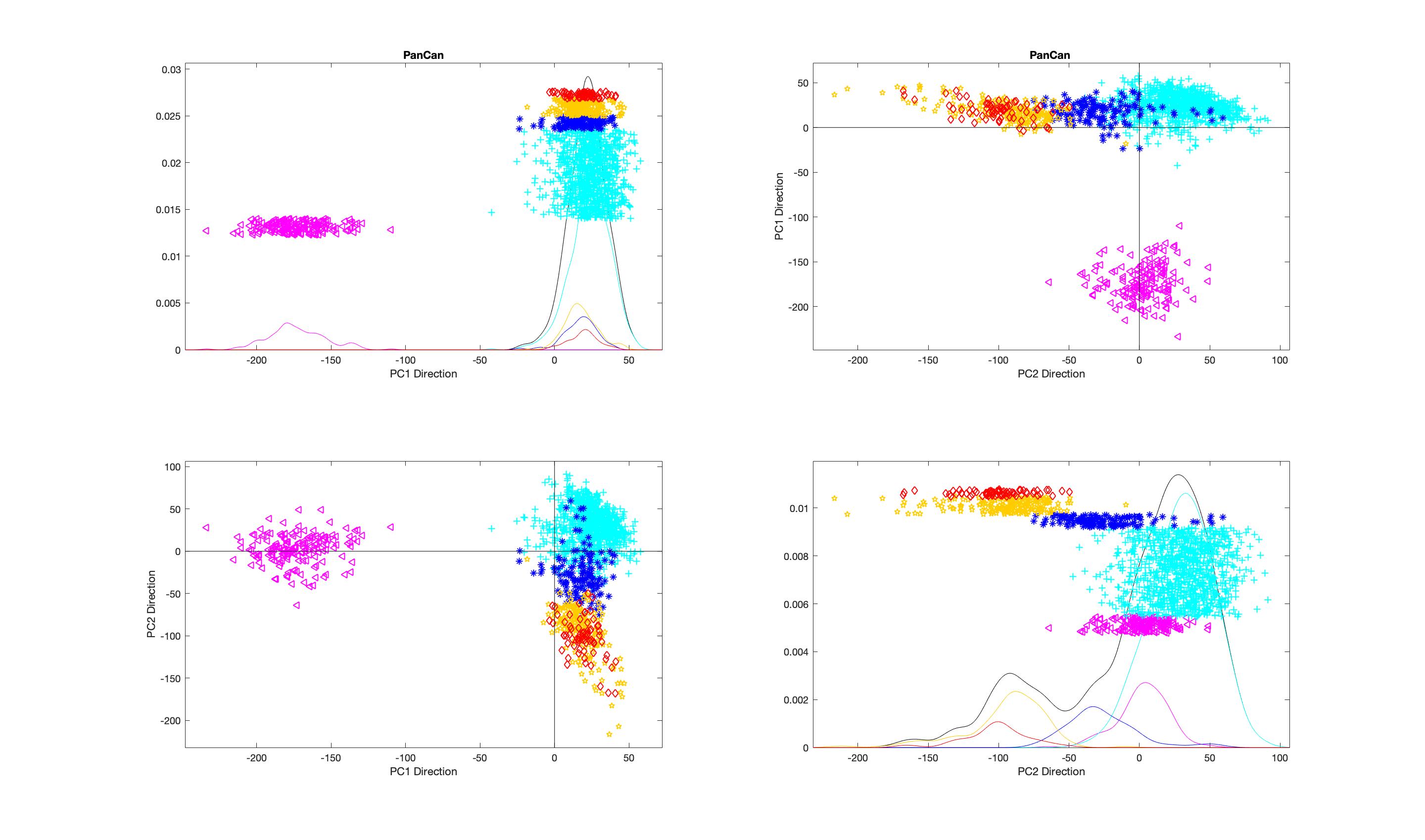}\\
  \caption{PCA scores scatter plot from TCGA Pan-Cancer gene expression data with symbols and colors in Table \ref{t1}. As biologically expected LAML is much different from the rest and COAD and READ overlap in PCA space.}
  \label{pca}
\end{figure}

Figure~\ref{pca} shows the relationship between cancer types using a PCA scatter plot, i.e., the two-dimensional projection of the point cloud in $\mathbb{R}^{12478}$ onto the two directions of highest variation. The PC1 scores are plotted on the vertical axis and the PC2 scores on the horizontal axis. The liquid tumor LAML is very different from the rest, which are solid epithelial tumors. Within the epithelial tumors, READ and COAD appear visually quite overlapped, consistent with the fact that these cells come from organs in the same developmental process and often referred to as a single disease (colorectal cancer). The BLCA and BRCA are somewhat different but not as separated as BLCA and LAML. 

\begin{figure}
  \centering
  \includegraphics[width=18cm]{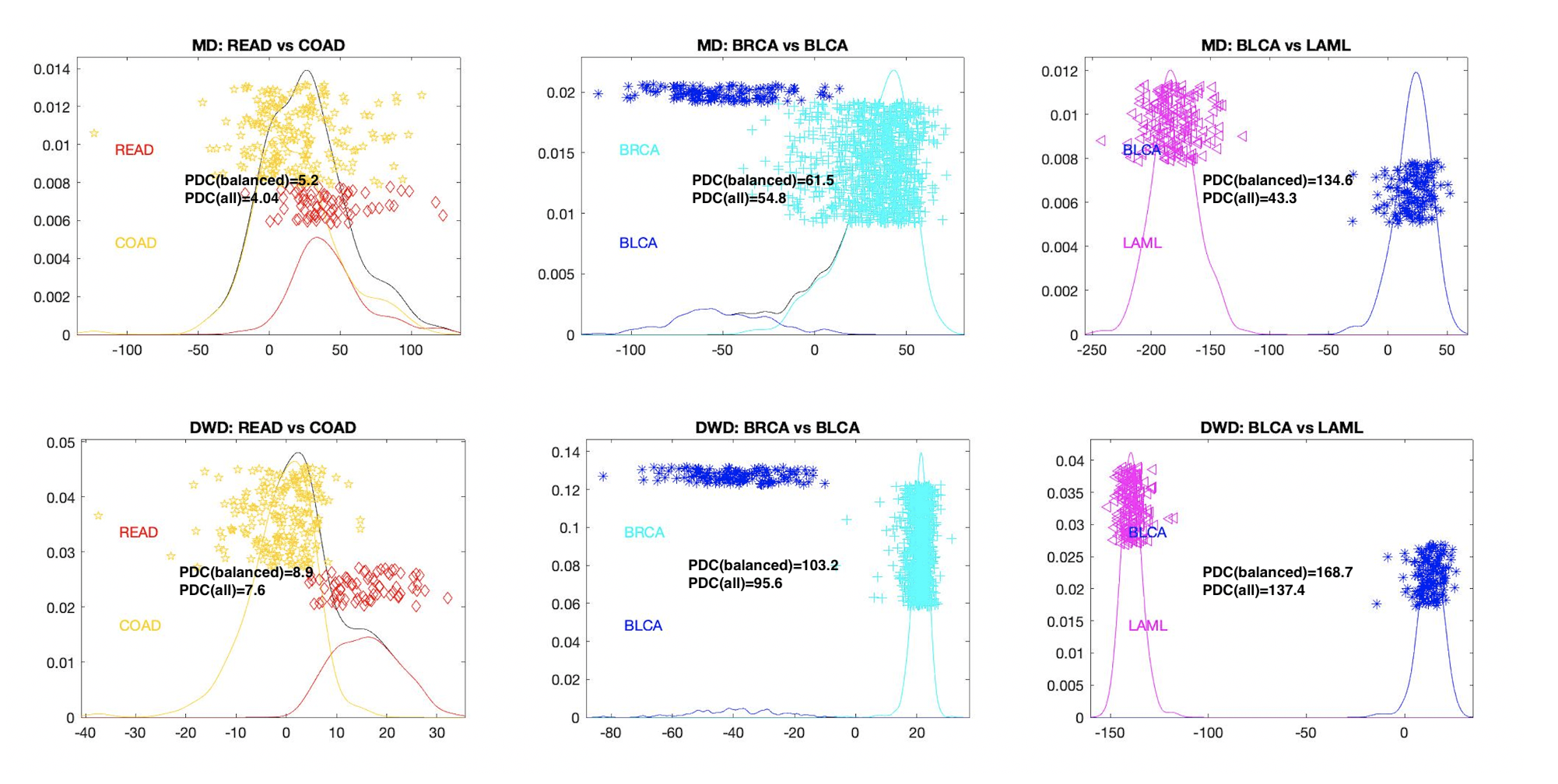}\\
  \caption{Distributions of projection scores on MD (top) and DWD (bottom) directions for the pairs: READ vs. COAD; BLCA vs. BRCA; LAML vs. BLCA. The $x$-axis is the value of projection scores, $y$-axis shows the height of the KDE estimates, and  
  the black text shows the corresponding PDCs. The separation of the subpopulations on the top (MD) is similar to those for the bottom (DWD) but not as distinct. PDCs for balanced permutations are higher than for all permutations.}
  \label{dwd}
\end{figure}

Figure~\ref{dwd} shows differences between three representative pairs of the subpopulations using projections on the MD (top row) and DWD (bottom row) directions. The subpopulations are indicated using symbols and colors described in Table~\ref{t1}. Each dot represents a projection of a case subject on the MD or DWD direction respectively. The value of the projection is displayed on the $x$-axis. The height of the dots are random for visual separation.  The curves are kernel density estimates of the projection scores with subdensities corresponding to the subpopulations. 
The $y$-axis shows the height of KDE densities. The black text shows the corresponding all and balanced permutation PDC respectively. As discussed in Section~\ref{Gaussian model simulation} there is better visual separation of the subpopulations for DWD (bottom) than for MD (top) and the PDCs for balanced permutations are higher than all permutations. This effect is particularly pronounced for the BLCA versus LAML, where the signal is the strongest. 
 Figure~\ref{dwd} shows broadly similar lessons to Figure~\ref{pca}: READ and COAD are rather overlapped (left panels); BRCA and BLCA are moderately different (middle panels) while LAML is very distinct from BLCA (right panels).

\begin{figure}[H]
  \centering
  \includegraphics[width=14cm]{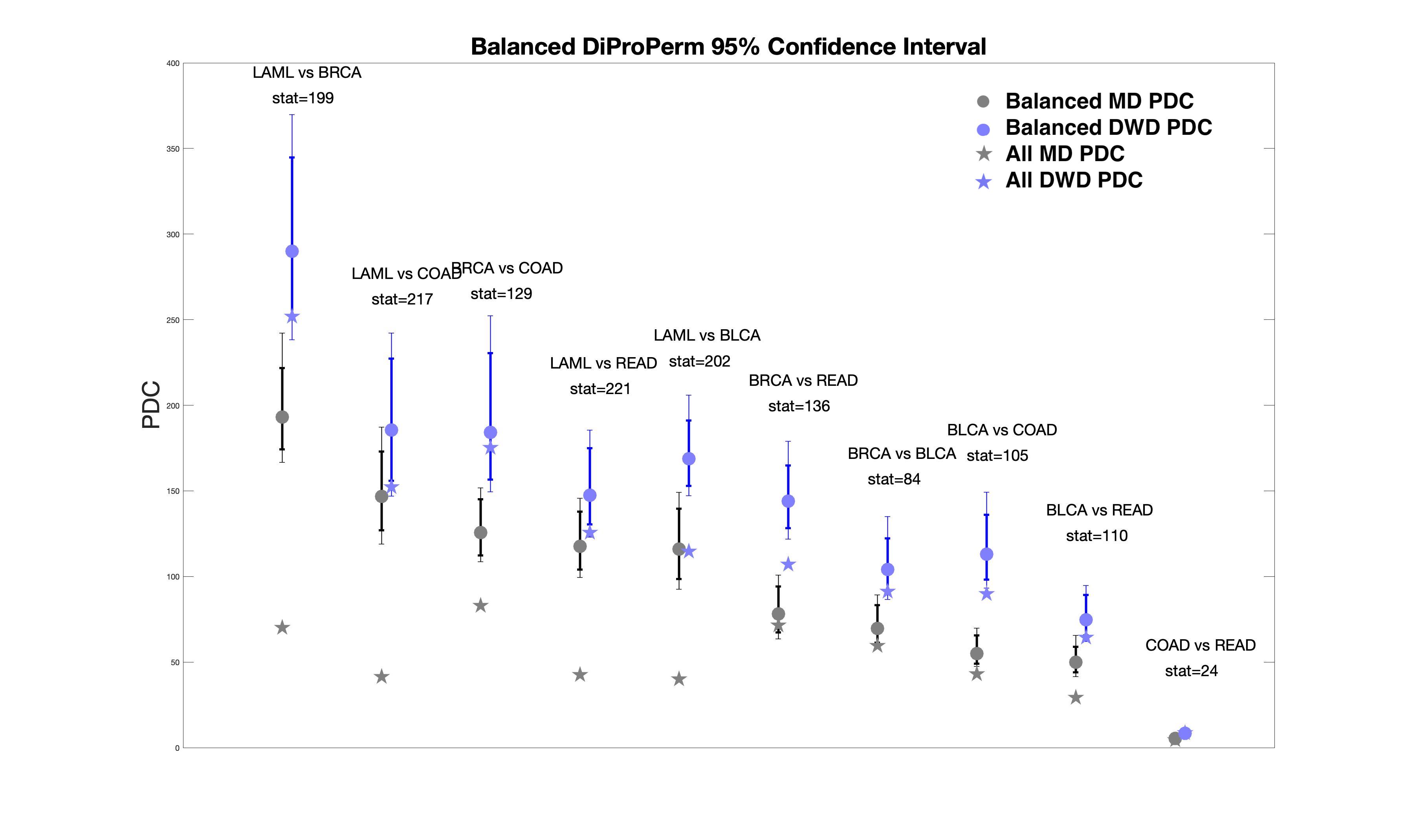}\\
 
  \caption{DiProPerm 95\% confidence intervals for all 10 pairwise tests of 5 types of cancers from TCGA data. The thicker lines represent individual confidence intervals for each PDC and the thinner lines are the  Bonferroni corrected confidence intervals. The circles and stars indicate the PDC estimates from balanced and all permutations respectively. The DWD-based PDCs are shown in blue/light blue and MD-based PDCs are shown in black/gray. The value of the test statistic is also printed at the top of each bar. This illustrates many effects discussed above.}
  \label{cip}
\end{figure}

For all 10 pairs of TCGA cancer types, Figure~\ref{cip} gives a comparison of the strength of separation using the PDC. The random permutation variability in estimating each PDC is reflected by a 95\% confidence interval as developed in Section~\ref{ci:all}. Conventional single-sample confidence intervals are shown as thick lines, the thin lines are Bonferroni-adjusted for the fact that we have 10 intervals. The results based on MD are shown in black/gray and the results based on DWD are shown in blue/light blue. The PDCs are not the centers of the confidence intervals because the distributions of the permutation statistics are skewed. The  PDCs computed using balanced permutations (circles) are much higher than the PDCs computed using all permutations (stars) showing the strong value of balanced permutations. Overall each DWD based PDC (blue) is higher than the corresponding MD based PDC (black) showing the utility of DWD over MD for distinguishing class differences in higher dimensions. 

The PDC allows us to accurately quantify the strength of population difference in each pair. The confidence intervals allow us to statistically compare these strengths of population difference across pairs. In Figure~\ref{cip}, all pairs involving LAML tend to have large PDCs, which is consistent with Figure~\ref{pca} which shows that LAML (magenta) is the most distinct cancer type. 
Pairs including BRCA also have relatively large PDCs. This is consistent with the fact that BRCA has the largest sample size which leads to a smaller variance and thus stronger statistical significance. The PDCs for LAML vs. READ and LAML vs. BLCA and BRCA vs. COAD reflect similar amounts of population difference. Those PDCs are the smallest among all test pairs indicating the weakest difference among the considered comparisons as shown in Figure~\ref{cip}.  This is consistent with the overlap of  COAD  and READ observed in Figures~\ref{pca} and \ref{dwd}.

In cases with a strong signal, such as LAML vs. BRCA, LAML vs. COAD and BRCA vs. COAD, the balanced PDCs (gray/blue circles) are much larger than the corresponding PDCs computed using all permutations (gray/blue stars). This is consistent with the idea that when the signal is strong, all permutations will cause a loss of power (see  Figure~\ref{81}). 
When the signal is weak, such as COAD vs. READ, all and the balanced PDCs are small and similar to each other. 
\section{Discussion}
\label{sb}
Our recommendation of balanced permutations is somewhat opposite to the recommendation against balanced permutations in \citet{southworth2009properties}. They appeal to group theory and suggest that all permutations are generally superior to balanced permutations since balanced permutations tend to be anti-conservative, i.e. their reported p-values are too small. In particular, under their null hypothesis, the permutation distribution, e.g., distribution of the red dots in Figure~\ref{3_densities} doesn't have enough extremely large values. \citet{hemerik2018exact} provided adjustments that make the use of balanced permutations for p-value calculation valid.  Appendix~\ref{a5} derives an often negligibly small alternative adjustment to both types of permutation PDC that overcomes the anti-conservative problem.

Figure~\ref{011} reveals the strange behavior that under the alternative the power of the tests from all permutations can decrease as the signal strength increases. The much-improved power of balanced permutations is shown in Figure~\ref{81} where the balanced permutation power as measured by PDC is proportional to the signal strength. When the signal is weak, Figure~\ref{81} shows that the balanced and all permutations give very similar PDCs. Thus balanced permutations are superior to all permutations in large-signal cases which often arise in bioinformatics and have no or minor differences from all permutations in small-signal cases.

 \section{Acknowledgement}
 We thank anonymous reviewers whose constructive comments helped us to greatly improve the presentation of our results. 
 Jan Hannig's research was supported in part by the National Science Foundation under Grant No. DMS-1916115, 2113404, and  2210337. J.S.~Marron's research was partially supported by the National Science Foundation Grant No. DMS-2113404.
 Katherine A. Hoadley's research was partially supported by Grant No. U24 CA264021.

\bibliographystyle{apalike}
\bibliography{ref}
\newpage
\begin{appendices}
\section{DWD optimization problem}
\label{ap:DWD}
The Distance Weighted Discrimination (DWD)
direction vector proposed by \citet{marron2007distance} as an improvement of the Support
Vector Machine (SVM) \citep{cortes1995support} classification method in high dimensions. In addition to improved classification
performance, DWD also provides unusually good visual separation of
subpopulations projected on the DWD direction because it avoids data piling issues (too many data points having a common projection) that is endemic to the SVM in high dimensions as discussed in \citet{marron2007distance}. 

The optimization problem of DWD (from \citet{marron2007distance}) is:
\begin{eqnarray*}
(P_{DWD}): \ \min_{\phi, \omega, \beta, \xi, \rho, ,\sigma, \tau} &\ &\{ Ce'\xi+e'\rho+e'\sigma\}\\
YX'\omega+\beta y+\xi-\rho+\sigma&=&0,\\
\phi&=&1,\\
\tau &=&e,\\
(\phi;\omega)\in S_{d+1},\  \xi\ge 0,\  (\rho_{i};\sigma_{i};\tau_{i})\in S_3,& & i=1, 2, ..., n.
\end{eqnarray*}
After some algebra, we obtain a simplified form of the dual as:
\begin{eqnarray*}
(D_{DWD})\ \max_{\alpha} -||XY\alpha||+2e'\sqrt{\alpha}, \quad y'\alpha=0,\ 0\le \alpha \le Ce
\end{eqnarray*}
(Here $\sqrt{\alpha}$ denotes the vector whose components are the square roots of those of $\alpha$).

\section{Conditional Distribution of  $\bar X_{per}- \bar Y_{per}$}
\label{a_condition}
Under the assumptions in Section \ref{cbp}, if we pick $r$ cases from each class to switch labels, 
\[\bar X_{per}|r\sim \frac{1}{m}N([r\times(-g)+(m-r)\times g]u, m\sigma^2I_d),\  \bar Y_{per}|r\sim \frac{1}{n}N([r\times g+(n-r)\times (-g)]u, n\sigma^2I_d) \]
thus
$$\bar X_{per}- \bar Y_{per}|r\sim N(2g(1-\frac{r}{m}-\frac{r}{n})u, \sigma^2(\frac{1}{m}+\frac{1}{n})I_d).$$

In this scheme, there are ${m\choose r} {n\choose n-r}$ permutations out of ${m+n\choose m}$ overall random permutations. Thus, the probability of picking $r$ cases from each class to switch labels is $\frac{{m\choose r} {n\choose n-r}}{{m+n\choose m}}$.

\section{Limit of PDC}
\label{ap:limit}
As shown in Section \ref{cbp}, $$E[C]=E[||\bar X-\bar Y||]=\sigma\sqrt{\frac{1}{m}+\frac{1}{n}}\sqrt{\frac{\pi}{2}}L_{1/2}^{d/2-1}(-\frac{2g^2}{\frac{1}{m}+\frac{1}{n}}),$$ so we need to calculate $E[C_i], Var[C_i]$ for both balanced and all permutations.

For balanced permutations, fix $r=\frac{mn}{m+n}$. It follows from \eqref{eq:mixture} that $C_i\sim \sigma\sqrt{\frac{1}{m}+\frac{1}{n}}\chi_d(0)$. Thus, 
\[
E[C_i]=\sigma\sqrt{\frac{1}{m}+\frac{1}{n}}\sqrt{2}\frac{\Gamma((d+1)/2)}{\Gamma(d/2)}, \ Var[C_i]=\sigma^2(\frac{1}{m}+\frac{1}{n})[d-2(\frac{\Gamma((d+1)/2)}{\Gamma(d/2)})^2]
\]
so 
\[f_b(m, n, d, g)=\frac{\sqrt{\frac{\pi}{2}}L_{1/2}^{d/2-1}(-\frac{2g^2}{\frac{1}{m}+\frac{1}{n}})-\sqrt{2}\frac{\Gamma((d+1)/2)}{\Gamma(d/2)}}{\sqrt{d-2(\frac{\Gamma((d+1)/2)}{\Gamma(d/2)})^2}}
\]
which is monotone increasing as a function of $g$ since aguerre polynomials are decreasing for negative values.

In the all permutation case, the limit (as $g\to \infty$) of the permuted MD direction converges to the observed MD direction. Consequently the limiting permutation distribution is the same for all $d$ as $g$ goes to infinity. This is consistent with what we observed in Figure~\ref{011}. 

For simplicity we investigate the case of $d=1$. We assume $\sigma=1$, then the $\chi_{1}(\lambda)$ is a \textit{folded normal distribution} (\citet{leone1961folded}) with the location parameter $\mu=\lambda$, and scale parameter $\sigma^2=1$.
Then 
\begin{eqnarray*}
E(C)&=&\sqrt{1/m+1/n}E(\chi_1(\sqrt{\frac{4g^2}{1/m+1/n}}))\\
&=&\sqrt{1/m+1/n}[\sqrt{2/\pi}\exp(-2g^2\frac{1}{1/m+1/n})+2g\sqrt{\frac{1}{1/m+1/n}}(1-2\Phi(-2g\sqrt{\frac{1}{1/m+1/n}}))].
\end{eqnarray*}
\begin{eqnarray*}
E(C_i)&=&\sqrt{1/m+1/n}E(\sum_{r=0}^m  p_R(r)\chi_1(\sqrt{\frac{4g_r^2}{1/m+1/n}}))\\
&=&\sqrt{1/m+1/n}[\sqrt{2/\pi}\sum_{r=0}^m p_R(r)\exp(-2g^2(1-r/m-r/n)^2\frac{1}{1/m+1/n})])\\
&+&2\sum_{r=0}^m  p_R(r) g|1-r/m-r/n| \sqrt{\frac{1}{1/m+1/n}}(1-2\Phi(-2g|1-r/m-r/n|\sqrt{\frac{1}{1/m+1/n}}))],\\
Var(C_i)&=&(1/m+1/n)(\sum_{r=0}^m  p_R(r) 4g^2(1-r/m-r/n)^2/(1/m+1/n)+1)-E(C_i)^2.
\end{eqnarray*}
Thus
\begin{eqnarray*}
\lim_{g\to \infty}\frac{E(C)-E(C_i)}{\sqrt{Var(C_i)}}=\lim_{g\to \infty}\frac{E(C)/g-E(C_i)/g}{\sqrt{Var(C_i)/g^2}}=\frac{\lim_{g\to \infty}E(C)/g-\lim_{g\to \infty}E(C_i)/g}{\lim_{g\to \infty}\sqrt{Var(C_i)/g^2}}.
\end{eqnarray*}
Since
\begin{eqnarray*}
\lim_{g\to \infty}E(C)/g&=&2\sqrt{1/m+1/n}\sqrt{\frac{1}{1/m+1/n}}=2\\
\lim_{g\to \infty}E(C_i)/g&=&2\sum_{r=0}^m  p_R(r)  |1-r/m-r/n|\\
\lim_{g\to \infty}\sqrt{Var(C_i)}/g&=&2\sqrt{1/m+1/n}\sqrt{\sum_{r=0}^m  p_R(r)  (1-r/m-r/n)^2-(\sum_{r=0}^m  p_R(r)  |1-r/m-r/n|)^2}\\
&=&2\sqrt{\frac{1}{m+n-1}-(\sum_{r=0}^m  p_R(r)  |1-r/m-r/n|)^2},
\end{eqnarray*}
for any $d$, the limiting value is:
\[
\lim_{g\to \infty}\frac{E(C)-E(C_i)}{\sqrt{Var(C_i)}}=\frac{1-\sum_{r=0}^m  p_R(r)  |1-r/m-r/n|}{\sqrt{\frac{1}{m+n-1}-(\sum_{r=0}^m  p_R(r)  |1-r/m-r/n|)^2}}.
\]

In our simulation, $m=n=100$, thus $\lim_{g\to \infty}(\text{PDC})\approx 23.48$, which is very close to the far right end of each (dashed/dotted) curve in Figure~\ref{011}. This indicates the high quality of this estimate of the limiting PDC.

\section{Permutation Correlation}
\label{a5}
Under the setting of Section \ref{cbp}, let us first consider $m=n$. As defined in Section \ref{Permutation distribution-f}, $R_i$ is the number of observations in each class that switch labels in permutation $i$. Let $\{X_{j}| j\in J_{R_i}\}$ be the cases in the original class +1 that are labeled as -1 and $\{Y_{k}| k\in K_{R_i}\}$ be the cases in the original class -1 that are labeled as +1 in one permutation.  Let $\sqrt{M_0}$ be the difference of the class means of the non-permuted data (let $R_0=0$) and let $\sqrt{M_i}$ be that of permutation $i$. Since 
$$Cov(M_0,M_i)=E(M_0M_i)-E(M_0)E(M_i),$$
we need $E(M_0)E(M_i)$ and $E(M_0M_i)$. From $M_0, M_i\sim \frac{2}{n}\sigma^2\chi^2_d(0)$, it is straightforward that 
$$E(M_0)E(M_i)=\frac{4d^2\sigma^4}{n^2}.$$
In order to calculate $E(M_0M_i)$, let
\begin{eqnarray*}
W&=&\frac{1}{n}(\sum_{k\in K_{R_i}}Y_k-\sum_{j\in J_{R_i}}X_j)\sim \sqrt{2R_i}\frac{\sigma}{n} N(0, 1)\\
U&=&\frac{1}{n}(\sum_{j\notin J_{R_i}}X_j-\sum_{k\notin K_{R_i}}Y_k)\sim \sqrt{2n-2R_i}\frac{\sigma}{n}N(0, 1)\\
M_0&=&||U-W||^2=||\frac{\sum X_j-\sum Y_k}{n}||^2\\
M_1&=&||U+W||^2=||\frac{\sum_{j \notin J_{R_i}} X_j-\sum_{j \in J_{R_i}} X_j-\sum_{k \notin K_{R_i}}Y_k+\sum_{k \in K_{R_i}}Y_k}{n}||^2
\end{eqnarray*}
and $W_j, U_j$ be entries of $W, U$. We have
\begin{eqnarray*}
E(M_0M_i)&=&\sum^d E[(U_j^4+W_j^4-2U_j^2W_j^2)]+\sum_{i\ne j}E[(U_i+W_i)^2(U_j-W_j)^2]\\
&=&\sum^d (E[(U_j^4]+E[W_j^4]-2E[U_j^2]E[W_j^2)])+\sum_{i\ne j}E[(U_i+W_i)^2]E[(U_j-W_j)^2]\\
&=&d[12R_i^2\frac{\sigma^4}{n^4}+12(n-R_i)^2\frac{\sigma^4}{n^4}-8R_i(n-R_i)\frac{\sigma^4}{n^4}]+4d(d-1)\frac{\sigma^4}{n^2}\\
&=&d\frac{\sigma^4}{n^4}[32R_i^2+12n^2-32R_in]+4d(d-1)\frac{\sigma^4}{n^2}.
\end{eqnarray*}
Thus
\begin{eqnarray*}
Cov(M_0, M_i)&=&d\frac{\sigma^4}{n^4}[32R_i^2+12n^2-32R_in]-4d\frac{\sigma^4}{n^2}\\
&=&d\frac{\sigma^4}{n^4}[32R_i^2+8n^2-32R_in]\\
&=&8d\frac{\sigma^4}{n^4}[4R_i^2+n^2-4R_in]\\
&=&8d\frac{\sigma^4}{n^4}(2R_i-n)^2
\end{eqnarray*}

As $d\to \infty$,
\begin{eqnarray*}
\sqrt{d}(\begin{bmatrix}
M_0/d \\
M_i/d
\end{bmatrix}-\begin{bmatrix}
\frac{2\sigma^2}{n} \\
\frac{2\sigma^2}{n} 
\end{bmatrix})&\xrightarrow[]{d}& N(0, \frac{8\sigma^4 }{n^2}\begin{bmatrix}
1& (\frac{2R_i-n}{n})^2\\
 (\frac{2R_i-n}{n})^2&1
\end{bmatrix})
\end{eqnarray*}
and by the delta method,
\begin{eqnarray*}
(\begin{bmatrix}
\sqrt{M_0} \\
\sqrt{M_i}
\end{bmatrix}-\begin{bmatrix}
\sqrt{\frac{2d\sigma^2}{n}} \\
\sqrt{\frac{2d\sigma^2}{n}} 
\end{bmatrix})&\xrightarrow[]{d}& N(0, \frac{\sigma^2 }{n}\begin{bmatrix}
1& (\frac{2R_i-n}{n})^2\\
 (\frac{2R_i-n}{n})^2&1
\end{bmatrix})
\end{eqnarray*}
so as $d\to \infty$, $Cor(\sqrt{M_0}, \sqrt{M_i})\to(\frac{2R_i-n}{n})^2$. 

In order to get the correlation between the mean differences from two random permutations, we add up the weighted correlations (the additive property applies due to this special setting, see the appendices for more details). For all permutations:
\begin{eqnarray*}
Corr_a&=&\frac{E[(\sqrt{M_0}-\mu_0)(\sqrt{M_i}-\mu_1)]}{\sqrt{Var(\sqrt{M_0})Var(\sqrt{M_i})}}=\frac{E[E[(\sqrt{M_0}-\mu_0)(\sqrt{M_i}-\mu_1)|R_i]]}{\sqrt{Var(\sqrt{M_0})Var(\sqrt{M_i})}}\\
&=&\frac{\sum_{R_i=0}^n \frac{{n \choose R_i}{n \choose n-R_i}}{{2n \choose n}}\frac{\sigma^2 }{n} (\frac{2R_i-n}{n})^2}{\sum_{R_i=0}^n \frac{{n \choose R_i}{n \choose n-R_i}}{{2n \choose n}}E[M_0]-(\sum_{R_i=0}^n \frac{{n \choose R_i}{n \choose n-R_i}}{{2n \choose n}}E[\sqrt{M_0}])^2}\\
&=&\frac{\sum_{R_i=0}^n \frac{{n \choose R_i}{n \choose n-R_i}}{{2n \choose n}}\frac{\sigma^2 }{n} (\frac{2R_i-n}{n})^2}{\frac{\sigma^2 }{n}}
=\sum_{R_i=0}^n \frac{{n \choose R_i}{n \choose n-R_i}}{{2n \choose n}}(\frac{2R_i-n}{n})^2=\frac{1}{2n-1}.
\end{eqnarray*}

\begin{figure}
  \centering
  \includegraphics[width=15cm]{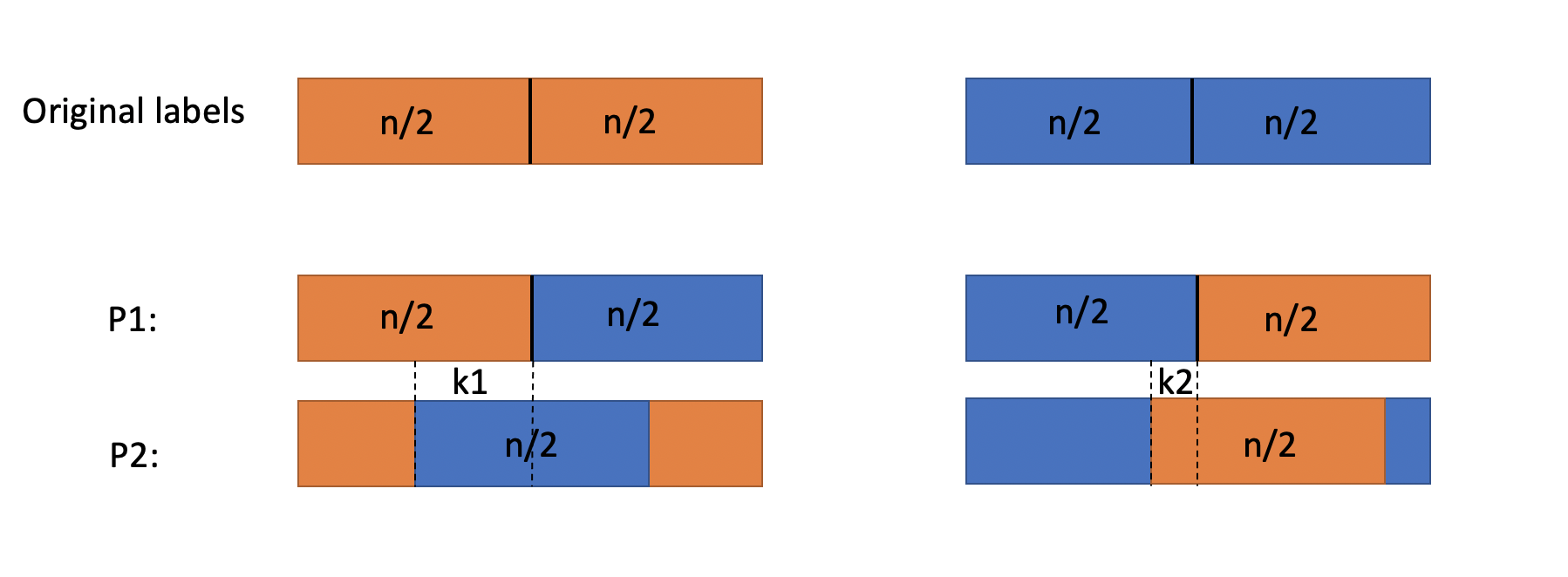}\\
  \caption{The first row represents the orginal class labels: orange vs. blue. The bottom two rows are two random balanced permutations. }
  \label{perp}
\end{figure}

For balanced permutations, we fix $R_i=\frac{n^2}{2n}=\frac{n}{2}$. In Figure~\ref{perp}, the first row represents the original labels and the bottom two rows represent two permutations denoted as P1 (P2). The orange and the blue represent 2 classes in each row. Covariance calculation in the case of balanced permutations is driven by two parameters $k1, k2$, which reflect the amount of overlap between P1 and P2. Let $\sqrt{M_1}$ be the distance between the two class means of P1 and let $\sqrt{M_2}$ be that of permutation P2. 
Let
\begin{eqnarray*}
\{\bar X_{per}-\bar Y_{per}\}_i&=&(X_{orange}+Y_{orange}-X_{blue}-Y_{blue})/n\sim \frac{1}{n}N(0, 2n\sigma^2), i=1, 2\\
U&=&p_1+p_2\sim \frac{1}{n}N(0, 2(n-k_1-k_2)\sigma^2)\\
W&=&p1-p2\sim \frac{1}{n}N(0, 2(k_1+k_2)\sigma^2)
\end{eqnarray*}
With a similar derivation, 
$$Cov(M_1, M_2)=8d\frac{\sigma^4}{n^4}(2(k_1+k_2)-n)^2$$

As $d\to \infty$, using a similar derivation as all permutations, the correlation between two random balanced permutations is:
$$Corr_b=\sum_{k_1=0}^{n/2} \sum_{k_2=0}^{n/2} \frac{{n/2 \choose k_1}{n/2 \choose n/2-k_1}}{{n \choose n/2}} \frac{{n/2 \choose k_2}{n/2 \choose n/2-k_2}}{{n \choose n/2}}(\frac{2(k_1+k_2)-n}{n})^2=\frac{1}{2n-2}.$$

\begin{figure}
  \centering
  \includegraphics[width=14cm]{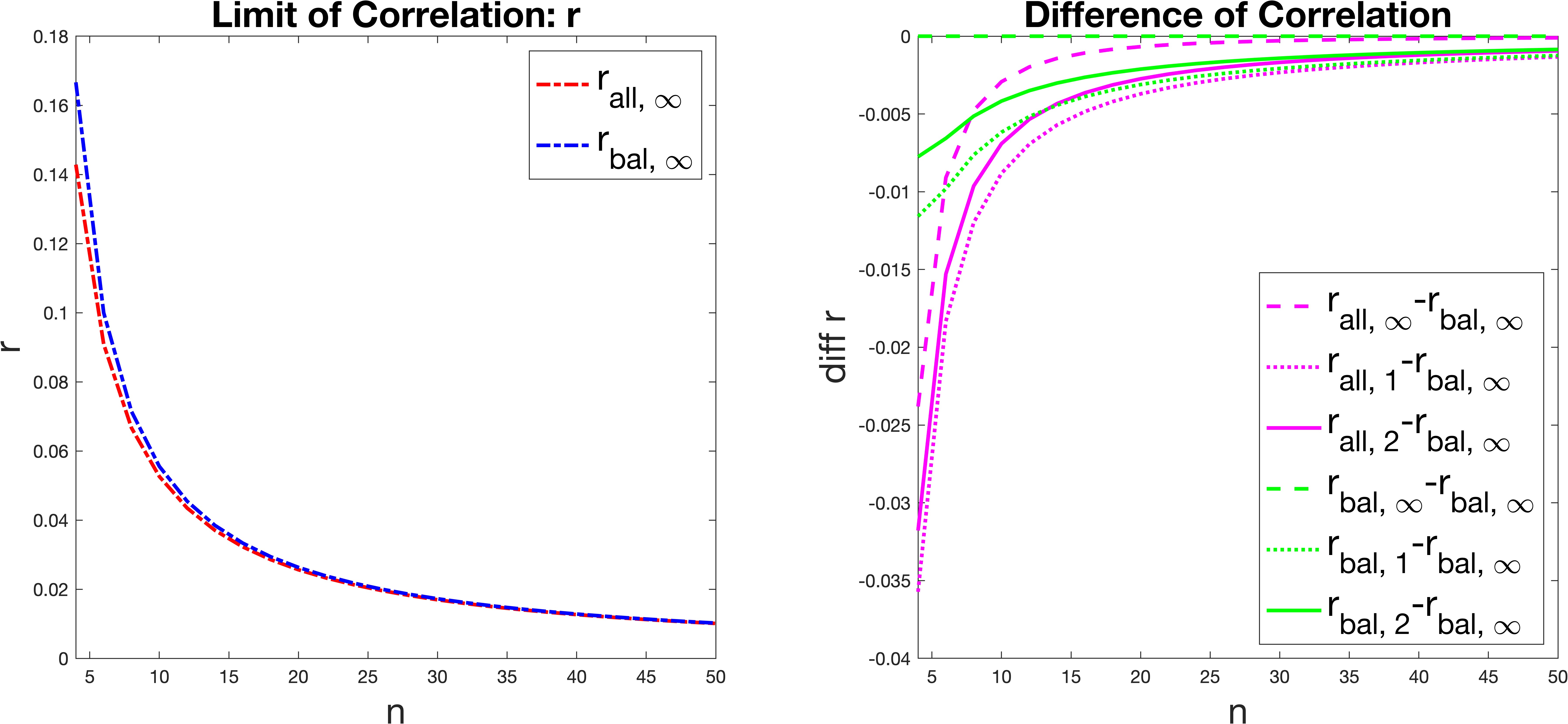}\\
  \caption{Correlation ($r$) as a function of $n$ with different line types and colors indicating different $d$ and balanced versus all permutations. In the left panel, when $d=\infty$, the red shows correlations of all permutations ($r_{all, \infty}$) with blue for balanced permutations ($r_{bal, \infty}$). These curves (decreasing rapidly) are very close to each other and $r_{bal, \infty}>r_{all, \infty}$. The right panel 
enables a more detailed study for $d=1,2$ in both cases, showing the difference between these correlations and the upper bound $r_{all, \infty}$. Correlations rapidly decrease as a function of $n$ and increase slightly as a function of $d$. When $d=1,2$, correlations are already very close to the limit $d=\infty$, which is also the upper bound. }
  \label{3}
\end{figure}

In the left panel of Figure~\ref{3}, we compare the theoretical correlations for balanced permutations (blue): $r_{bal, \infty}=\frac{1}{2n-2}$ and for all permutations (red): $r_{all, \infty}=\frac{1}{2n-1}$ as functions of $n$. The blue is always larger than the red showing that the balanced permutations have larger correlations than all permutations. Those correlations are all very small and decrease rapidly with $n$. When $n>50$, the difference between them is quite negligible and the correlations will decrease as $n$ goes to infinity with a limit of zero. 

The right panel of Figure~\ref{3} is a zoomed in view to investigate the difference between correlations for $d=1,2,\infty$. The curves are all differences of correlations from balanced permutations when $d=\infty$, i.e. differences from $r_{bal, \infty}$. All curves are smaller than or equal to zero indicating that $r_{bal, \infty}$ is the correlation upper bound. In particular, Figure~\ref{3} shows that $r_{bal, \infty}\ge r_{bal, 2}\ge r_{bal, 1}$ and $r_{bal, \infty}\ge r_{all, \infty}\ge r_{all, 2}\ge r_{all, 1}$. 

The dotted magenta curve is the smallest which indicates that the correlation for all permutations with $d=1$ ($r_{all,1}$) is the smallest. For each color, magenta, and green, the solid curves are larger than the dotted and the dashed are larger than the solid. This suggests that correlations will increase when $d$ increases for both balanced and all permutations. As $n$ becomes large, all curves are close to each other, indicating the correlations all become similar and very small. 

In the case of $m\ne n$, similar results hold $Corr_a$.  In particular, similar calculations show $Corr_b$ are close to each other, but the all permutations correlation is always less than that for balanced permutations:
\begin{enumerate}
\item $Corr_a=\frac{1}{2(\frac{2}{\frac{1}{m}+\frac{1}{n}})-1}=\frac{m+n}{4mn-m-n}$
\item $Corr_b=\frac{1}{2(\frac{2}{\frac{1}{m}+\frac{1}{n}})-2}=\frac{m+n}{4mn-2m-2n}$
\end{enumerate}

The sample variance $S^2$, correlation and variance are related by: $$E(S^2)=Var\times(1-Corr).$$ Thus instead of the original DiProPerm PDC: $\frac{C-\bar C}{S}$, we propose $\frac{(C-\bar C)\sqrt{1-Corr}}{S}$ as the adjusted PDC. From here on, we will use the adjusted PDC in this paper. This adjustment makes very little difference unless the sample size is very small. 



\end{appendices}
\end{document}